\documentclass[12pt,prd,showpacs,tightenlines,nofootinbib]{revtex4}
\usepackage{bm}
\usepackage{graphics}
\usepackage{rotating}
\usepackage{epsfig}
\begin{document}
\title{
Semileptonic decays of  $\Lambda_c$ baryons in
the relativistic quark model}  
\author{R. N. Faustov}
\author{V. O. Galkin}
\affiliation{Institute of Informatics in Education, FRC CSC RAS,
  Vavilov Street 40, 119333 Moscow, Russia}

\begin{abstract}
Motivated by recent experimental progress in studying weak decays of
the $\Lambda_c$ baryon we investigate its semileptonic decays in the
framework of the relativistic quark model based on the quasipotential
approach with the QCD-motivated potential. The form factors of the $\Lambda_c\to \Lambda l\nu_l$ and $\Lambda_c\to nl\nu_l$ decays are calculated in the whole accessible kinematical region without extrapolations and additional model assumptions. Relativistic effects are systematically taken into account including transformations of baryon wave functions from the rest to moving reference frame and  contributions of the intermediate negative-energy states. Baryon wave functions found in the previous mass spectrum calculations are used for the numerical evaluation. Comprehensive predictions for decay rates, asymmetries and polarization parameters are given. They agree well with available experimental data.   
\end{abstract}

\pacs{13.30.Ce, 12.39.Ki, 14.20.Mr, 14.20.Lq}

\maketitle

\section{Introduction}

Recently significant experimental progress has been achieved in
studying weak decays of the charm baryons. Thus in 2014, Belle
Collaboration \cite{belle} measured the branching fraction
$Br(\Lambda_c^+\to pK^-\pi^+)= (6.84 \pm 0.24^{+0.21}_{-0.27})\%$ with
the larger value and precision improved by a factor of 5 over previous
results \cite{cleo}. This measurement is very important since many of
the previously measured $\Lambda_c$ branching fractions are determined
by their ratio with respect to $Br(\Lambda_c^+\to pK^-\pi^+)$
\cite{rosner}. It leads to the improved value of the semileptonic
branching fraction $Br(\Lambda_c^+\to \Lambda e^+\nu_e)=(2.9\pm0.5)\%$
\cite{pdg}. Last year the  BESIII Collaboration \cite{besiii} reported
the first absolute measurement of the branching fraction of the
semileptonic $\Lambda_c^+\to \Lambda e^+\nu_e$ process which is considerably
more precise than previous values \cite{pdg}. The branching fraction
was found to be $Br(\Lambda_c^+\to \Lambda
e^+\nu_e)=(3.63\pm0.38\pm0.20)\%$ \cite{besiii}.  Later BESIII
provided measurements for twelve $\Lambda_c^+$ decay modes, including
the branching fraction for $\Lambda_c^+ \to pK^-\pi^+$ decay with the value $(5.84 \pm 0.27 \pm 0.23)\%$ \cite{besiii2}. The measurements of the branching fractions of the other Cabibbo-Kobayashi-Maskawa (CKM) favored \cite{besiii2} and singly suppressed \cite{besiii3} hadronic decay modes were significantly improved.

Motivated by these experimental achievements we investigate
semileptonic $\Lambda_c$ decays in the relativistic quark model based
on the quasipotential approach with the QCD-motivated potential.  The mass spectra of heavy and
strange baryons calculated in the quark-diquark picture of baryons in
our model \cite{barregge} agree well with available experimental data.
The proton and neutron masses are also well reproduced. The study of
baryon spectroscopy allowed us to determine the baryon wave functions
which we use here for the numerical calculations. Recently we
considered the semileptonic $\Lambda_b$ decays \cite{lbdecay}. General
relativistic expressions for the decay form factors as overlap
integrals of the initial and final baryon wave functions were
found. They are obtained without application of either heavy quark
$1/m_Q$ or nonrelativistic $v/c$ expansions. Relativistic effects,
including wave function transformations from the rest to moving reference frame and contributions of the intermediate negative-energy states are consistently taken into account. It is important to emphasize that the momentum transfer $q^2$ dependence of the decay form factors is explicitly 
determined without additional model assumptions and
extrapolations. Here we apply these expressions for the calculation of
form factors of the $\Lambda_c$ semileptonic decays. On their basis we
present predictions for differential and total branching fractions of
the $\Lambda_c\to \Lambda l\nu_l$ and $\Lambda_c\to nl\nu_l$ decays ($l=e,\mu$) and their asymmetries and polarization parameters.

\section{Relativistic quark model}

In the relativistic quark model based on the quark-diquark picture and
the quasipotential approach the interaction of two quarks in a diquark and the quark-diquark interaction  in a baryon are described by the
diquark wave function $\Psi_{d}$ of the bound quark-quark state
and by the baryon wave function $\Psi_{B}$ of the bound quark-diquark
state,  which satisfy the relativistic
quasipotential equation of the Schr\"odinger type \cite{mass}
\begin{equation}
\label{quas}
{\left(\frac{b^2(M)}{2\mu_{R}}-\frac{{\bf
p}^2}{2\mu_{R}}\right)\Psi_{d,B}({\bf p})} =\int\frac{d^3 q}{(2\pi)^3}
 V({\bf p,q};M)\Psi_{d,B}({\bf q}),
\end{equation}
where the relativistic reduced mass and the center-of-mass system
relative momentum squared on mass shell are
\[
\mu_{R}=\frac{M^4-(m^2_1-m^2_2)^2}{4M^3}, \qquad{b^2(M) }
=\frac{[M^2-(m_1+m_2)^2][M^2-(m_1-m_2)^2]}{4M^2},
\]
and $M$ is the bound state mass (diquark or baryon),
$m_{1,2}$ are the masses of  quarks ($q_1$ and $q_2$) which form
the diquark or of the  diquark ($d$) and  quark ($q$) which form
the baryon ($B$), and ${\bf p}$  is their relative momentum.

The quasipotentials $V({\bf p,q};M)$  of
the quark-quark or quark-diquark interaction are constructed with
the help of the QCD-motivated off-mass-shell scattering amplitude, projected onto the positive
energy states. The effective quark interaction is taken to be the
sum of the usual one-gluon exchange term and the mixture of long-range
vector and scalar linear confining potentials with the mixing
coefficient $\varepsilon$. It is also assumed that
the vector confining potential contains not only the Dirac term but
the Pauli term, thus introducing the anomalous chromomagnetic quark moment $\kappa$. The explicit expressions for the
quasipotentials  are given in Ref.~\cite{hbar}. 

In the nonrelativistic limit  the
usual Cornell-like potential is reproduced
\begin{equation}
V(r)=-\frac43\frac{\alpha_s}{r}+Ar+B,
\end{equation}
where the QCD coupling constant with freezing is given by
\begin{equation}
  \label{eq:alpha}
  \alpha_s(\mu^2)=\frac{4\pi}{\displaystyle\left(11-\frac23n_f\right)
\ln\frac{\mu^2+M_B^2}{\Lambda^2}}, 
\qquad \mu=\frac{2m_1m_2}{m_1+m_2},
\end{equation} 
$n_f$ is the number of flavors, and the background mass $M_B=2.24\sqrt{A}=0.95$~GeV,
$\Lambda=413$~MeV \cite{ltetr}.

All parameters of the model were
fixed previously from calculations of meson and baryon properties
\cite{mass,hbar}. The constituent quark masses $m_u=m_d=0.33$
GeV, $m_s=0.5$ GeV, $m_c=1.55$ GeV  and the parameters of the linear potential
$A=0.18$ GeV$^2$ and $B=-0.3$ GeV have the usual values of quark
models.  The value of the mixing coefficient of vector and scalar
confining potentials $\varepsilon=-1$ has been determined from the
consideration of the heavy quark expansion for the semileptonic heavy
meson decays and charmonium radiative decays \cite{mass}. The
universal Pauli interaction constant $\kappa=-1$  has been fixed from
the analysis of the fine splitting of heavy quarkonia ${}^3P_J$-
states \cite{mass}.  Note that the long-range chromomagnetic
contribution to the potential, which is proportional to $(1+\kappa)$,
vanishes for the chosen value of $\kappa=-1$.

\section{Semileptonic decay form factors}

To calculate the heavy $\Lambda_c$ baryon decay rate to the  $\Lambda$ hyperon or
neutron ($n$) it is necessary to determine the
corresponding matrix element of the  weak current between baryon
states. In the quasipotential approach it is expressed by the relation  
\begin{equation}\label{mxet} 
\langle \Lambda(n)(p_{{\Lambda(n)}}) \vert J^W_\mu \vert \Lambda_c(p_{\Lambda_c})\rangle
=\int \frac{d^3p\, d^3q}{(2\pi )^6} \bar \Psi_{\Lambda(n)\,{\bf p}_{{\Lambda(n)}}}({\bf
p})\Gamma _\mu ({\bf p},{\bf q})\Psi_{\Lambda_c\,{\bf p}_{\Lambda_c}}({\bf q}),
\end{equation}
where $\Gamma _\mu ({\bf p},{\bf
q})$ is the two-particle vertex function and  
$\Psi_{B\,{\bf p}_{B}}$ are the $B$ ($B=\Lambda_c,\Lambda,n$) 
baryon  wave functions projected onto the positive-energy states of 
quarks. The vertex function $\Gamma$ receives
relativistic contributions both from the impulse approximation diagram
and from the
diagrams with the intermediate negative-energy states which are the consequence
of the projection onto the positive-energy states in the
quasipotential approach \cite{lbdecay}. The boosts of the $B$ baryon wave
functions from the rest to the moving reference frame with momentum ${\bf p}_{B}$ are also
consistently taken into account \cite{lbdecay}.

The hadronic matrix elements for the semileptonic decay $\Lambda_c\to
\Lambda(n)l\nu_l$  can be parameterized  in terms of six invariant form factors:
\begin{eqnarray}
  \label{eq:llff}
  \langle \Lambda(n)(p',s')|V^\mu|\Lambda_c(p,s)\rangle&=& \bar
  u_{\Lambda(n)}(p',s')\Bigl[F_1(q^2)\gamma^\mu+F_2(q^2)\frac{p^\mu}{M_{\Lambda_c}}+F_3(q^2)\frac{p'^\mu}{M_{\Lambda(n)}}\Bigl]
u_{\Lambda_c}(p,s),\cr
 \langle \Lambda(n)(p',s')|A^\mu|\Lambda_c(p,s)\rangle&=& \bar
  u_{\Lambda(n)}(p',s')\Bigl[G_1(q^2)\gamma^\mu+G_2(q^2)\frac{p^\mu}{M_{\Lambda_c}}+G_3(q^2)\frac{p'^\mu}{M_{\Lambda(n)}}\Bigl]
\gamma_5 u_{\Lambda_c}(p,s),\qquad 
\end{eqnarray}
where   $u_{\Lambda_{c}}(p,s)$ and
$u_{\Lambda(n)}(p',s')$ are Dirac spinors of the initial and final
baryon; $q=p'-p$.
In the literature another parametrization is often employed \cite{giklsh,gikls}
\begin{eqnarray}
  \label{eq:ff}
  \langle \Lambda(n)(p',s')|V^\mu|\Lambda_c(p,s)\rangle&=& \bar
  u_{\Lambda(n)}(p',s')\Bigl[f_1^V(q^2)\gamma^\mu-f_2^V(q^2)i\sigma^{\mu\nu}\frac{q_\nu}{M_{\Lambda_c}}+f_3^V(q^2)\frac{q^\mu}{M_{\Lambda_c}}\Bigl]
u_{\Lambda_c}(p,s),\cr
 \langle \Lambda(n)(p',s')|A^\mu|\Lambda_c(p,s)\rangle&=& \bar
  u_{\Lambda(n)}(p',s')[f_1^A(q^2)\gamma^\mu-f_2^A(q^2)i\sigma^{\mu\nu}\frac{q_\nu}{M_{\Lambda_c}}+f_3^A(q^2)\frac{q^\mu}{M_{\Lambda_c}}\Bigl]
\gamma_5 u_{\Lambda_c}(p,s),\qquad 
\end{eqnarray}
Relations between the two sets of form factors are given in Ref. \cite{lbdecay}.

Expressions for the form factors $F_i(q^2)$, $G_i(q^2)$ ($i=1,2,3$)
obtained in our model are given in Ref. \cite{lbdecay}. They are valid
both for the heavy-to-heavy and heavy-to-light baryon decays. The form factors are
expressed through the overlap integrals of the baryon wave
functions which we take from the mass spectrum calculations. All
relativistic effects including transformations of the baryon wave
functions from the rest to moving reference frame and contributions of
the intermediate negative-energy states are consistently taken into account. It is
important to point out that the momentum transfer $q^2$ behavior is
explicitly determined in the whole kinematical range without
extrapolations or model assumptions which are used in most of other
theoretical considerations. This fact improves reliability of the form
factor calculations. Note that in the heavy quark limit these form
factors satisfy model independent relations imposed by the heavy quark
symmetry \cite{iwb}.

Our numerical analysis shows that the weak decay form factors can 
be approximated with good accuracy in the physical region [$0\le
q^2\le q^2_{\rm max}=(M_{\Lambda_c}-M_{\Lambda(n)})^2$] by the following expression: 
\begin{equation}
  \label{fitff}
  F(q^2)=\frac{F(0)}{\displaystyle\left(1-\sigma_1\frac{q^2}{M^2_{\Lambda_c}}+\sigma_2  \frac{q^4}{M_{\Lambda_c}^4}+
      \sigma_3\frac{q^6}{M_{\Lambda_c}^6}+
      \sigma_4\frac{q^8}{M_{\Lambda_c}^8}\right)}.
\end{equation}
The difference between the fitted and numerical values of the form
factors is less than 0.5\%.  The analytical properties of these form factors are discussed in Ref.\cite{wsdm}.

\begin{figure}
\centering
  \includegraphics[width=8cm]{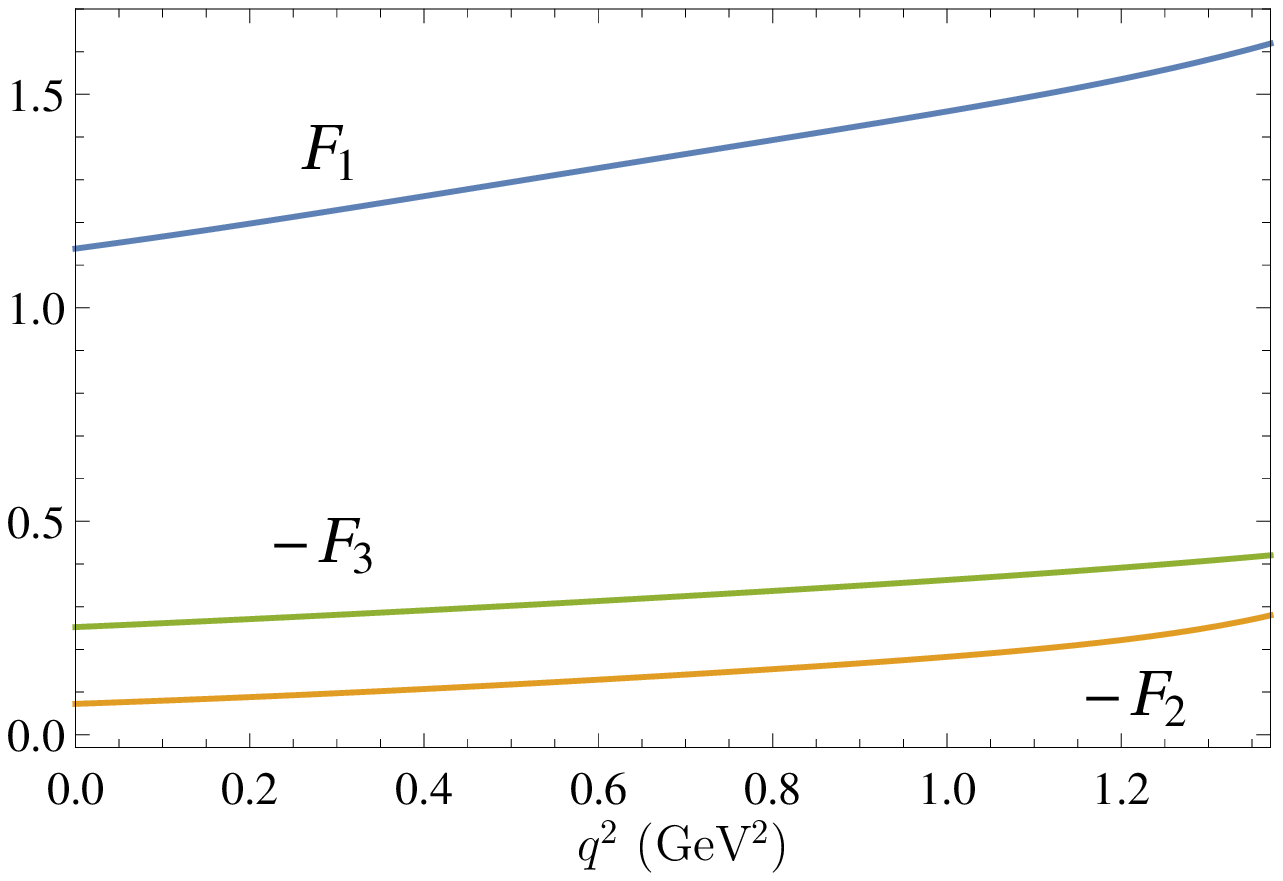}\ \
 \ \includegraphics[width=8cm]{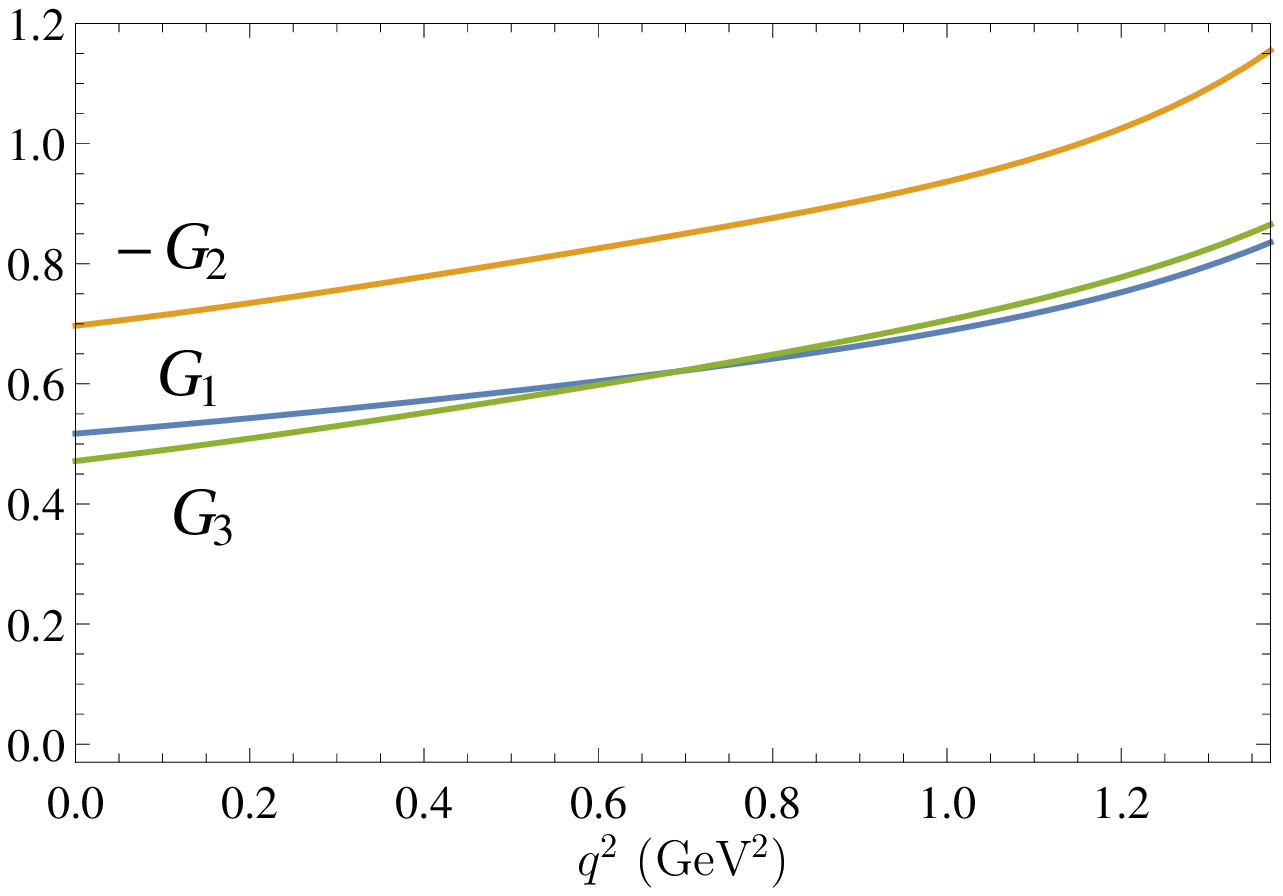}\\
\caption{Form factors of the weak $\Lambda_c\to \Lambda$ transition.    } 
\label{fig:ffLcL}
\end{figure}

\begin{figure}
\centering
  \includegraphics[width=8cm]{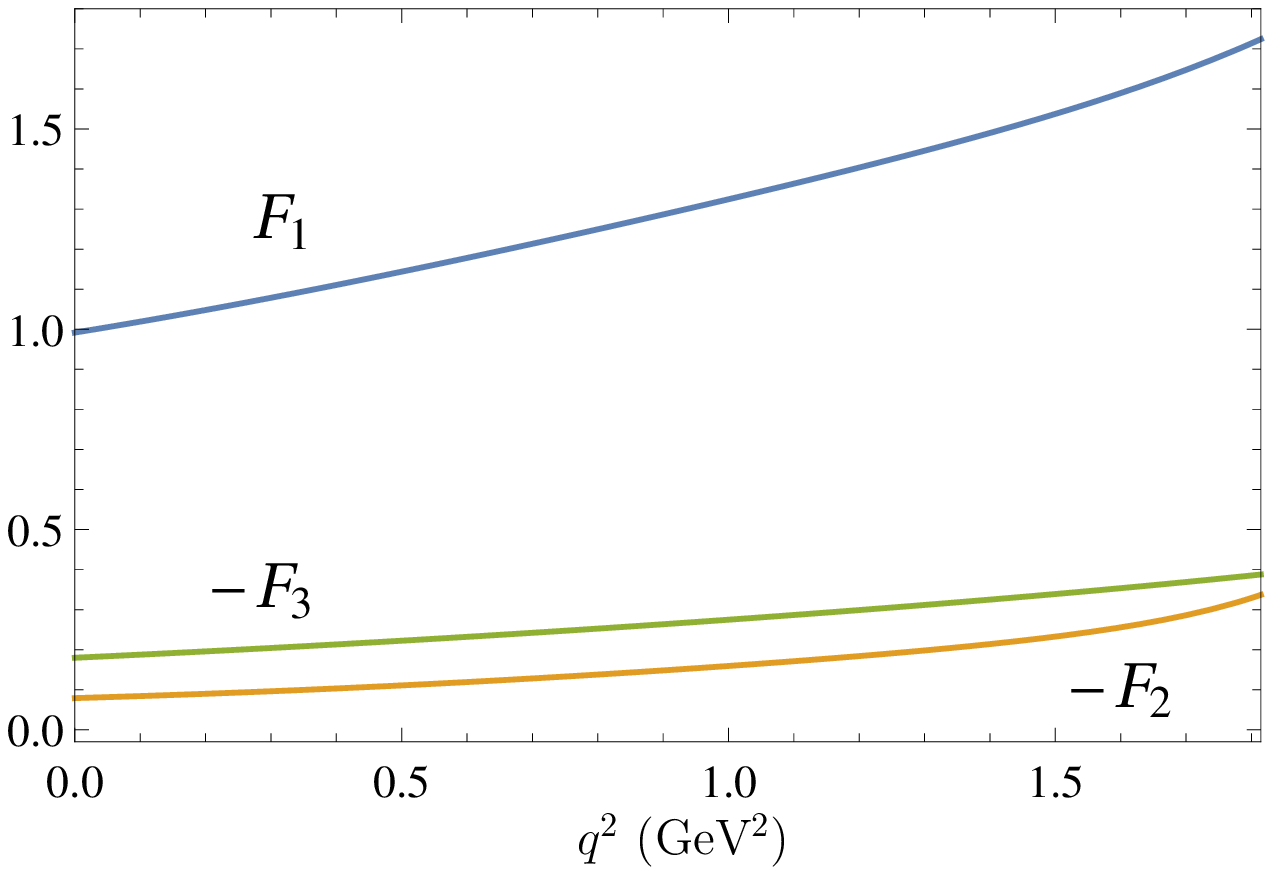}\ \
 \ \includegraphics[width=8cm]{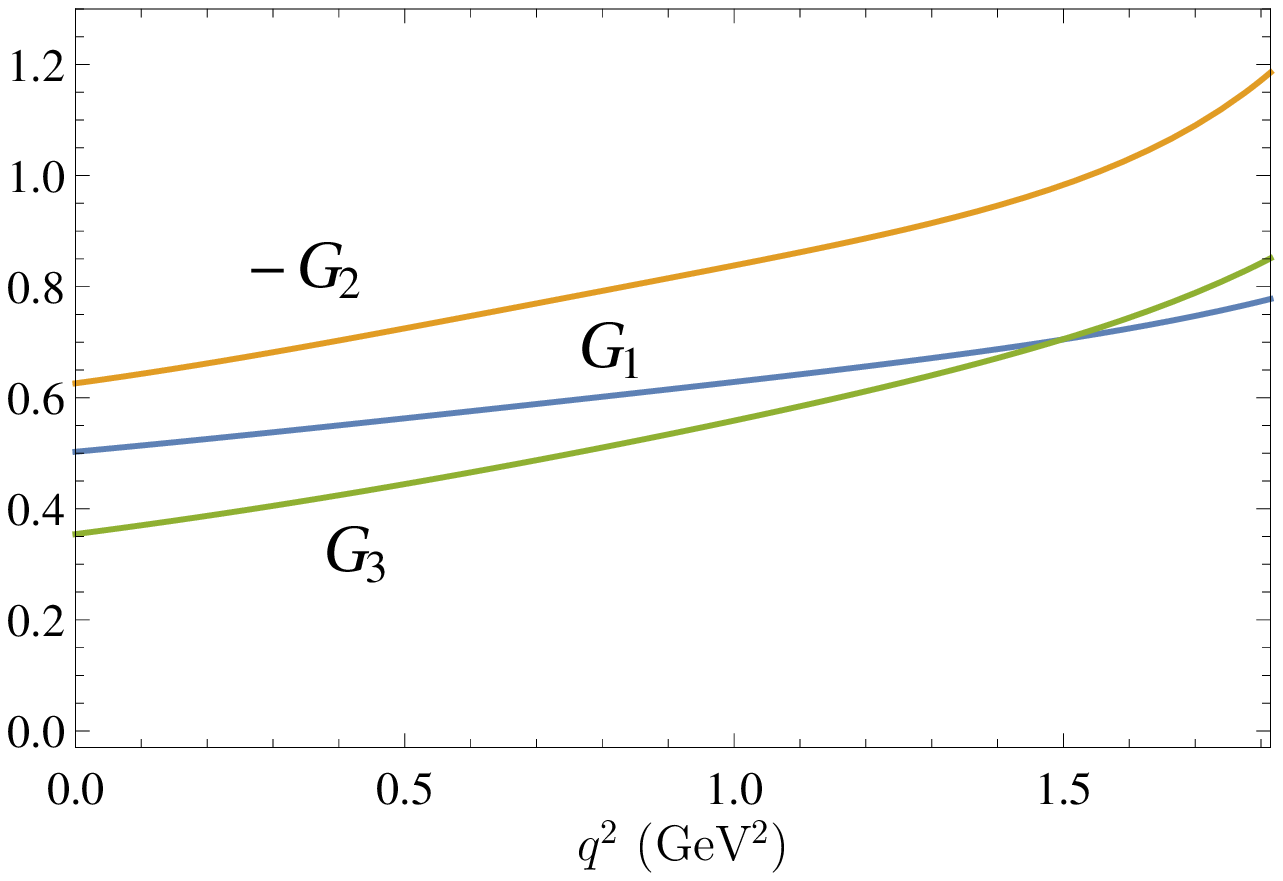}\\
\caption{Form factors of the weak $\Lambda_c\to n$ transition.    } 
\label{fig:ffLcn}
\end{figure}

The values of the form factor parameters $F(0)$
and $\sigma_{1,2,3,4}$, as well as the values at zero recoil  $F(q^2_{\rm max})$, are given in
Tables~\ref{ffLcL},~\ref{ffLcn}. We can estimate the errors of the
form factor calculations only within our model, since the uncertainty of the model itself is  unknown. They  mostly originate form the
uncertainties in the baryon wave functions and  from the subleading
contributions in the low recoil region. The latter one is suppressed
since the subleading contributions are proportional to the ratio of the
small binding energy to the baryon mass.
We estimate them to be less than 5\%.  The $\Lambda_c$ baryon decay form factors
are plotted in Figs.~\ref{fig:ffLcL} and \ref{fig:ffLcn}.

\begin{table}
\caption{Form factors of the weak $\Lambda_c\to \Lambda$ transition. }
\label{ffLcL}
\begin{ruledtabular}
\begin{tabular}{ccccccc}
& $F_1(q^2)$ & $F_2(q^2)$& $F_3(q^2)$& $G_1(q^2)$ & $G_2(q^2)$ &$G_3(q^2)$\\
\hline
$F(0)$          &1.14 &$-0.072$ & $-0.252$ & 0.517 & $-0.697$&0.471\\
$F(q^2_{\rm max})$&1.62  &$-0.280$ & $-0.420$ & 0.836& $-1.16$& 0.865\\
$\sigma_1$      &$1.24$&$5.15$& $1.85$& $1.20$ &$1.23$&  1.91\\
$\sigma_2$      &$-1.58$&$12.1$&$1.03$&$-1.33$&$-4.07$&$-1.18$\\
$\sigma_3$      &$19.3$&$1.03$& $4.31$& $12.1$ &$37.1$&  21.5\\
$\sigma_2$      &$-44.4$&$-51.2$&$-13.2$&$-40.7$&$-97.6$&$-54.9$\\
\end{tabular}
\end{ruledtabular}
\end{table}

\begin{table}
\caption{Form factors of the weak $\Lambda_c\to n$ transition. }
\label{ffLcn}
\begin{ruledtabular}
\begin{tabular}{ccccccc}
& $F_1(q^2)$ & $F_2(q^2)$& $F_3(q^2)$& $G_1(q^2)$ & $G_2(q^2)$ &$G_3(q^2)$\\
\hline
$F(0)$          &0.992 &$-0.079$ & $-0.180$ & 0.503 & $-0.626$&0.354\\
$F(q^2_{\rm max})$&1.72  &$-0.337$ & $-0.388$ & 0.778& $-1.19$& 0.851\\
$\sigma_1$      &$1.39$&$3.16$& $2.21$& $1.14$ &$1.34$&  2.26\\
$\sigma_2$      &$-0.683$&$0.029$&$2.17$&$-0.264$&$-2.68$&$0.800$\\
$\sigma_3$      &$8.43$&$22.4$& $-0.039$& $6.77$ &$23.1$&  8.84\\
$\sigma_2$      &$-14.7$&$-41.7$&$-2.09$&$-14.4$&$-44.7$&$-18.3$\\
\end{tabular}
\end{ruledtabular}
\end{table}

We compare our results for the form factors $f^{V,A}_{1,2,3}$  at the
maximum recoil point $q^2=0$ with the predictions of other approaches
in Table~\ref{compbpiff}. The covariant confined  quark model was used
in Ref.~\cite{gikls}. The authors of Refs.~\cite{lhw,kkmw}
employ QCD light-cone sum rules. The calculations in
Ref. \cite{abss} are based on  full QCD sum rules at light
cone. We find reasonable agreement with results of
Refs. \cite{gikls,lhw,kkmw}, while the predictions of
Ref. \cite{abss} are substantially different for most of the form factors.

\begin{table}
\caption{Theoretical predictions for the form factors of 
  weak baryon decays at maximum
  recoil point $q^2=0$.  }
\label{compbpiff}
\begin{ruledtabular}
\begin{tabular}{ccccccc}
&$f^V_1(0)$&$f^V_2(0)$&$f^V_3(0)$&$f^A_1(0)$&$f^A_2(0)$&$f^A_3(0)$\\
\hline
$\Lambda_c\to\Lambda$\\
this paper& 0.700& 0.295& 0.222& 0.448& $-0.135$& $-0.832$\\
\cite{gikls}&0.511&0.289& $-0.014$& 0.466& $-0.025$&$-0.400$\\
\cite{lhw} &0.517&0.123& &0.517& $-0.123$\\
$\Lambda_c\to n$\\
this paper& 0.627& 0.259& 0.179& 0.433& $-0.118$& $-0.744$\\
\cite{gikls}&0.470&0.246& $0.039$& 0.414& $-0.073$&$-0.328$\\
\cite{kkmw}&$0.59^{+0.15}_{-0.16}$&$0.43^{+0.13}_{-0.12}$&
&$0.55^{+0.14}_{-0.15}$ &$-0.16^{+0.08}_{-0.05}$&\\
\cite{abss}&0.17&1.78&2.95&0.52&0.71&$-0.0073$\\
\end{tabular}
\end{ruledtabular}
\end{table}

\section{Semileptonic decay observables}

Now we use the calculated $\Lambda_c$ baryon form factors for the
evaluation of the semileptonic decay rates, polarization observables
and decay asymmetries. For this purpose it is convenient to employ the
helicity formalism  \cite{kk}. In it the helicity amplitudes are
expressed in terms of the form factors \cite{kk} by the following relations
\begin{eqnarray}
  \label{eq:ha}
  H^{V,A}_{+1/2,\, 0}&=&\frac1{\sqrt{q^2}}{\sqrt{2M_{\Lambda_c}M_{\Lambda(n)}(w\mp 1)}}
[(M_{\Lambda_c} \pm M_{\Lambda(n)}){\cal F}^{V,A}_1(w) \pm M_{\Lambda(n)}
(w\pm 1){\cal F}^{V,A}_2(w)\cr
&& \pm M_{\Lambda_{c}} (w\pm 1){\cal F}^{V,A}_3(w)],\cr
 H^{V,A}_{+1/2,\, 1}&=&-2\sqrt{M_{\Lambda_c}M_{\Lambda(n)}(w\mp 1)}
 {\cal F}^{V,A}_1(w),\cr
 H^{V,A}_{+1/2,\, t}&=&\frac1{\sqrt{q^2}}{\sqrt{2M_{\Lambda_c}M_{\Lambda(n)}(w\pm 1)}}
[(M_{\Lambda_c} \mp M_{\Lambda(n)}){\cal F}^{V,A}_1(w) \pm(M_{\Lambda_c}- M_{\Lambda(n)}w
){\cal F}^{V,A}_2(w)\cr
&& \pm (M_{\Lambda_{c}} w- M_{\Lambda(n)}){\cal F}^{V,A}_3(w)],
\end{eqnarray}
where 
$$w=\frac{M_{\Lambda_c}^2+M_{\Lambda(n)}^2-q^2}
{2M_{\Lambda_c}M_{\Lambda(n)}},$$ 
 the upper (lower)  sign corresponds  to $V(A)$ and ${\cal F}^V_i\equiv F_i$,
${\cal F}^A_i\equiv G_i$ ($i=1,2,3$). $H^{V,A}_{\lambda',\,
  \lambda_W}$ are the helicity  
amplitudes for weak transitions induced by vector ($V$) and axial
vector ($A$) currents, where $\lambda'$ and $\lambda_W$ are the
helicities of the final baryon and the virtual $W$-boson, respectively. 
The amplitudes for negative values of the helicities are related to
the ones with the positive values by
$$H^{V,A}_{-\lambda',\,-\lambda_W}=\pm H^{V,A}_{\lambda',\, \lambda_W}.$$
The total helicity amplitude for the
$V-A$ current can be written as
$$H_{\lambda',\, \lambda_W}=H^{V}_{\lambda',\, \lambda_W}
-H^{A}_{\lambda',\, \lambda_W}.$$
It is convenient to introduce the following set of helicity structure
functions \cite{giklsh,gikls}
\begin{eqnarray}
  \label{eq:hhc}
  {\cal H}_U&=&|H_{+1/2,+1}|^2+|H_{-1/2,-1}|^2,\cr
{\cal H}_L&=&|H_{+1/2,0}|^2+|H_{-1/2,0}|^2,\cr
{\cal H}_S&=&|H_{+1/2,t}|^2+|H_{-1/2,t}|^2,\cr
  {\cal H}_P&=&|H_{+1/2,+1}|^2-|H_{-1/2,-1}|^2,\cr
{\cal H}_{L_P}&=&|H_{+1/2,0}|^2-|H_{-1/2,0}|^2,\cr
{\cal H}_{S_P}&=&|H_{+1/2,t}|^2-|H_{-1/2,t}|^2.
\end{eqnarray}
Then the differential decay rate for the semileptonic $\Lambda_c$ baryon
decay to the $\Lambda(n)$ is given by \cite{gikls}
\begin{equation}
  \label{eq:dgamma}
  \frac{d\Gamma(\Lambda_c\to \Lambda(n)l\nu_l)}{dq^2}=\frac{G_F^2}{(2\pi)^3}
  |V_{cq}|^2
  \frac{\lambda^{1/2}(q^2-m_l^2)^2}{48M_{\Lambda_c}^3q^2}{\cal H}_{tot},
\end{equation}
where $G_F$ is the Fermi constant, $V_{cq}$ ($q=s,d$) is the CKM matrix element, $\lambda\equiv
\lambda(M_{\Lambda_c}^2,M_{\Lambda(n)}^2,q^2)=M_{\Lambda_c}^4+M_{\Lambda(n)}^4+q^4-2(M_{\Lambda_c}^2M_{\Lambda(n)}^2+M_{\Lambda(n)}^2q^2+M_{\Lambda_c}^2q^2)$,
$m_l$ is the lepton mass and
\begin{eqnarray}
  \label{eq:hh}
  {\cal H}_{tot}&=&({\cal H}_U+{\cal H}_L) \left(1+\frac{m_l^2}{2q^2}\right)+\frac{3m_l^2}{2q^2}{\cal H}_S .
\end{eqnarray}
The other useful observables for the semileptonic $\Lambda_c$ decays are the following.\\ 
a) Forward-backward asymmetry of the charged lepton
\begin{eqnarray}
  \label{eq:afb}
  A_{FB}(q^2)&=&\frac{\frac{d\Gamma}{dq^2}({\rm forward})-\frac{d\Gamma}{dq^2}({\rm backward})}{\frac{d\Gamma}{dq^2}}\cr
&=&\frac34\frac{{\cal H}_P-2\frac{m_l^2}{q^2}(H_{+1/2,0}H_{+1/2,t}^\dag+H_{-1/2,0}H_{-1/2,t}^\dag)}{{\cal H}_{tot}}.\qquad
\end{eqnarray}
b) The convexity parameter
\begin{equation}
  \label{eq:cf}
  C_F(q^2)=\frac34\left(1-\frac{m_l^2}{q^2}\right)\frac{{\cal H}_U-2{\cal H}_L}{{\cal H}_{tot}}.
\end{equation}
c) The longitudinal polarization of the final $\Lambda(n)$ baryon
\begin{equation}
\label{eq:pl}
P_L(q^2)=\frac{({\cal H}_P+{\cal
    H}_{L_P})\left(1+\frac{m_l^2}{2q^2}\right)+3
  \frac{m_l^2}{2q^2}{\cal H}_{S_P}}{{\cal H}_{tot}}.
\end{equation}
The forward-backward asymmetry of the charged lepton $A_{FB}$ and the
convexity parameter $C_F$ are the linear and quadratic in $\cos\theta$
terms of the twofold angular distribution $d\Gamma/d q^2d\cos\theta$ for the decay $\Lambda_c\to
\Lambda(n)W^+(\to l^+\nu_l)$, where $\theta$ is the angle between
lepton and $W$ \cite{giklsh}. Detailed experimental study of such
 distribution can lead to the measurement of their values. On the
other hand, the longitudinal polarization of the final baryon can be
extracted from the experimental study of the fourfold angular
distributions $\Lambda_c\to\Lambda(\to p\pi^-)W^+(\to l^+\nu_l)$
\cite{giklsh,gikls}. Such complicated analysis can in principle be
done by the BESIII Collaboration.    

\begin{figure}
  \centering
 \includegraphics[width=8cm]{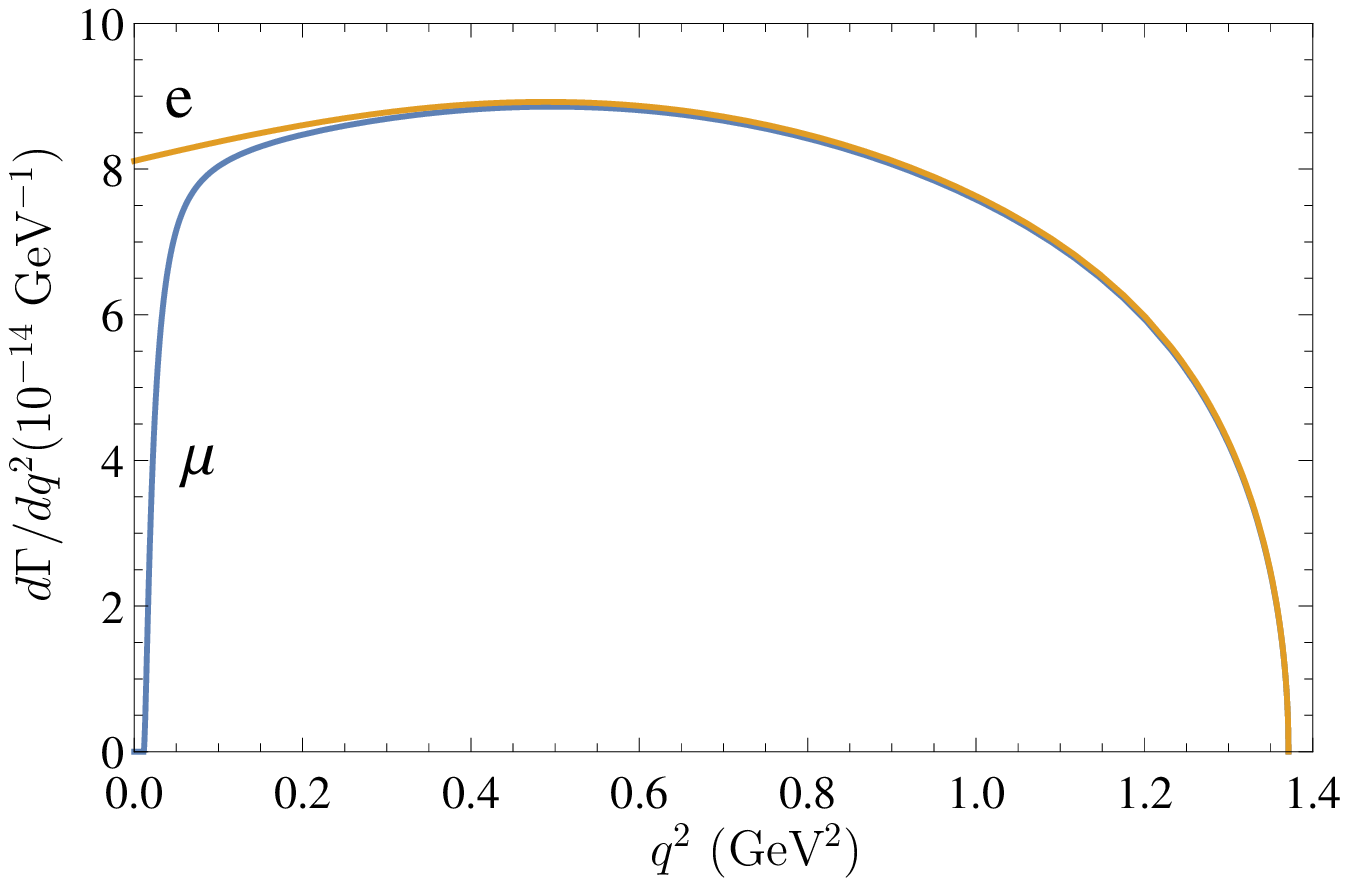}\ \
 \  \includegraphics[width=8cm]{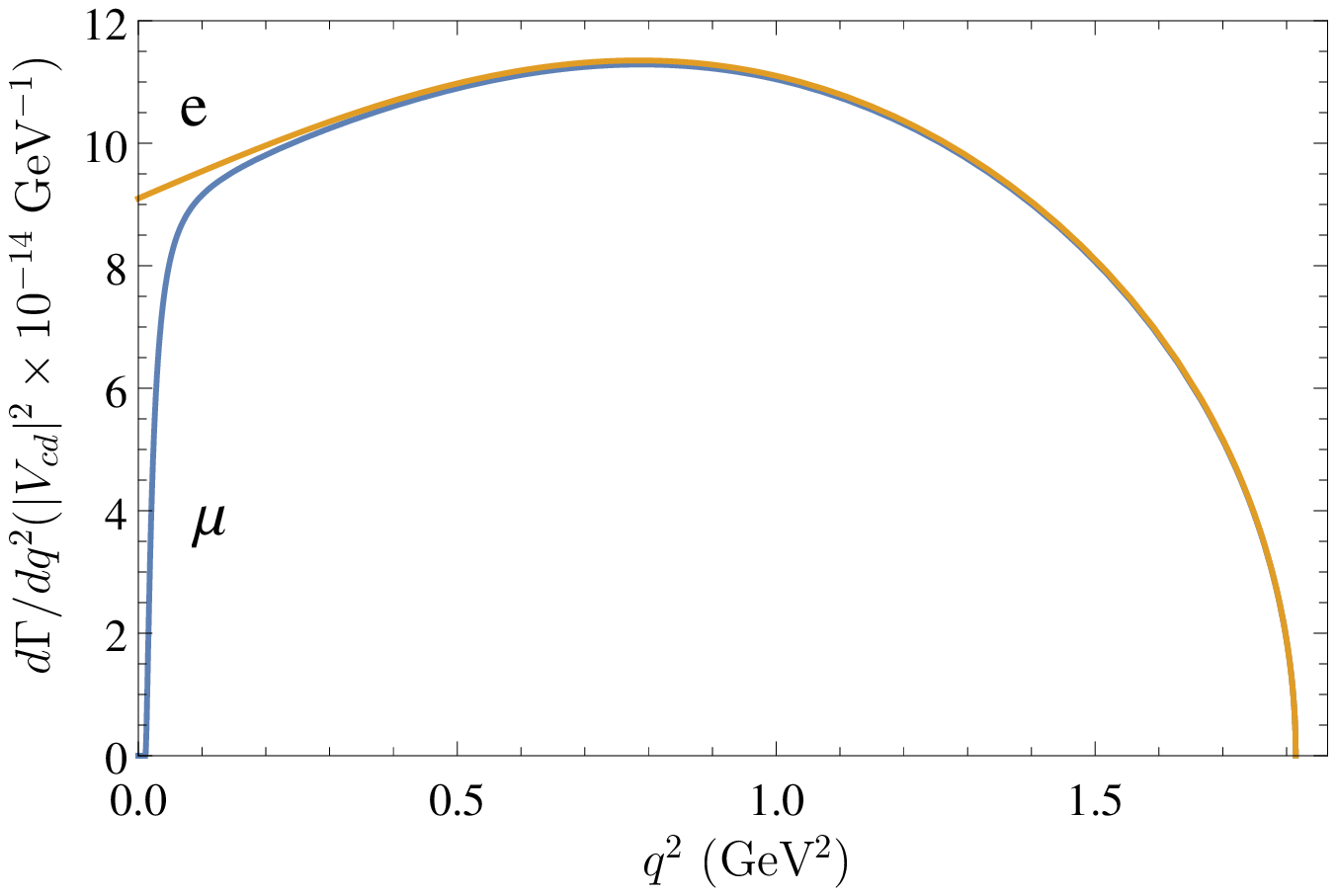}

  \caption{Differential decay rates  of the
    $\Lambda_c^+\to \Lambda l^+\nu_l$ (left) and $\Lambda_c^+\to n l^+\nu_l$ (right)
    semileptonic decays. }
  \label{fig:brLc}
\end{figure}

\begin{figure}
  \centering
 \includegraphics[width=8cm]{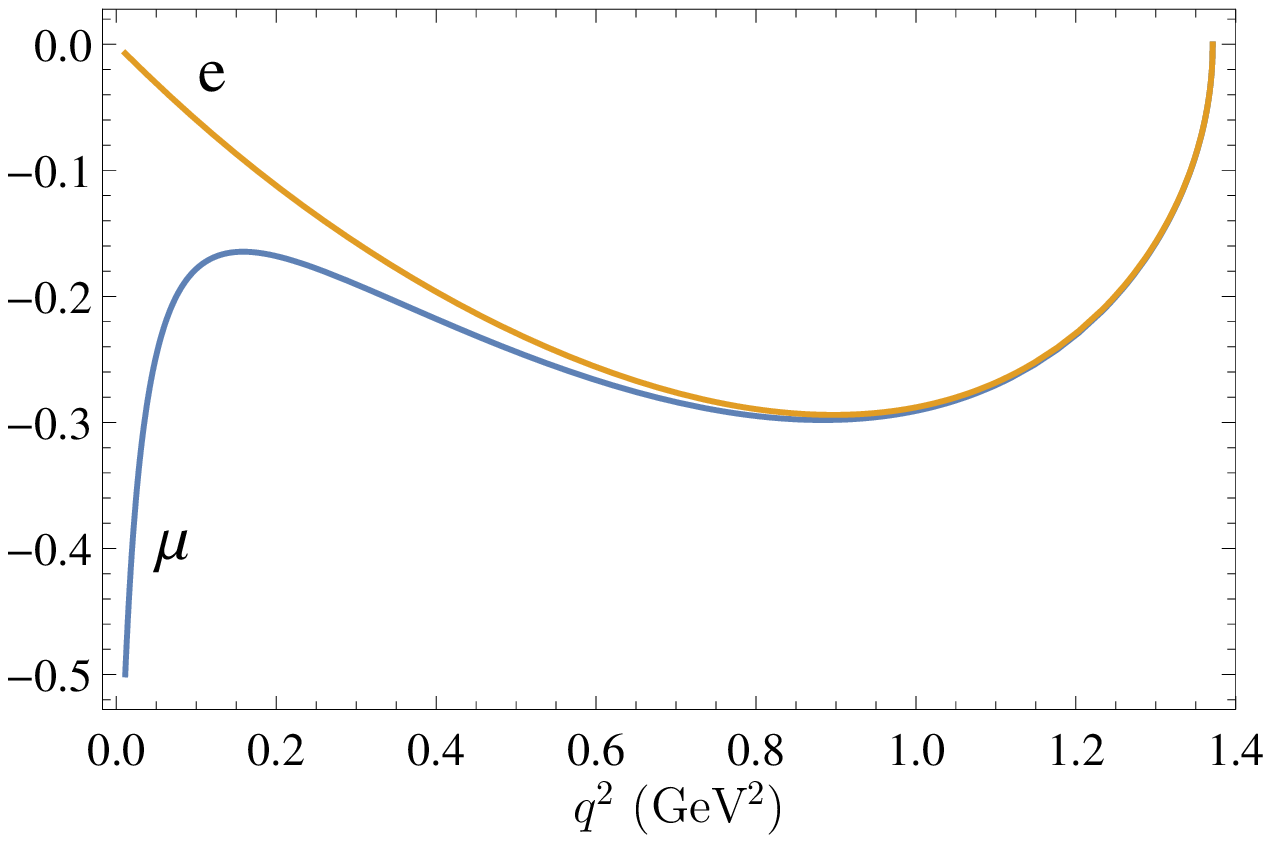}\ \
 \  \includegraphics[width=8cm]{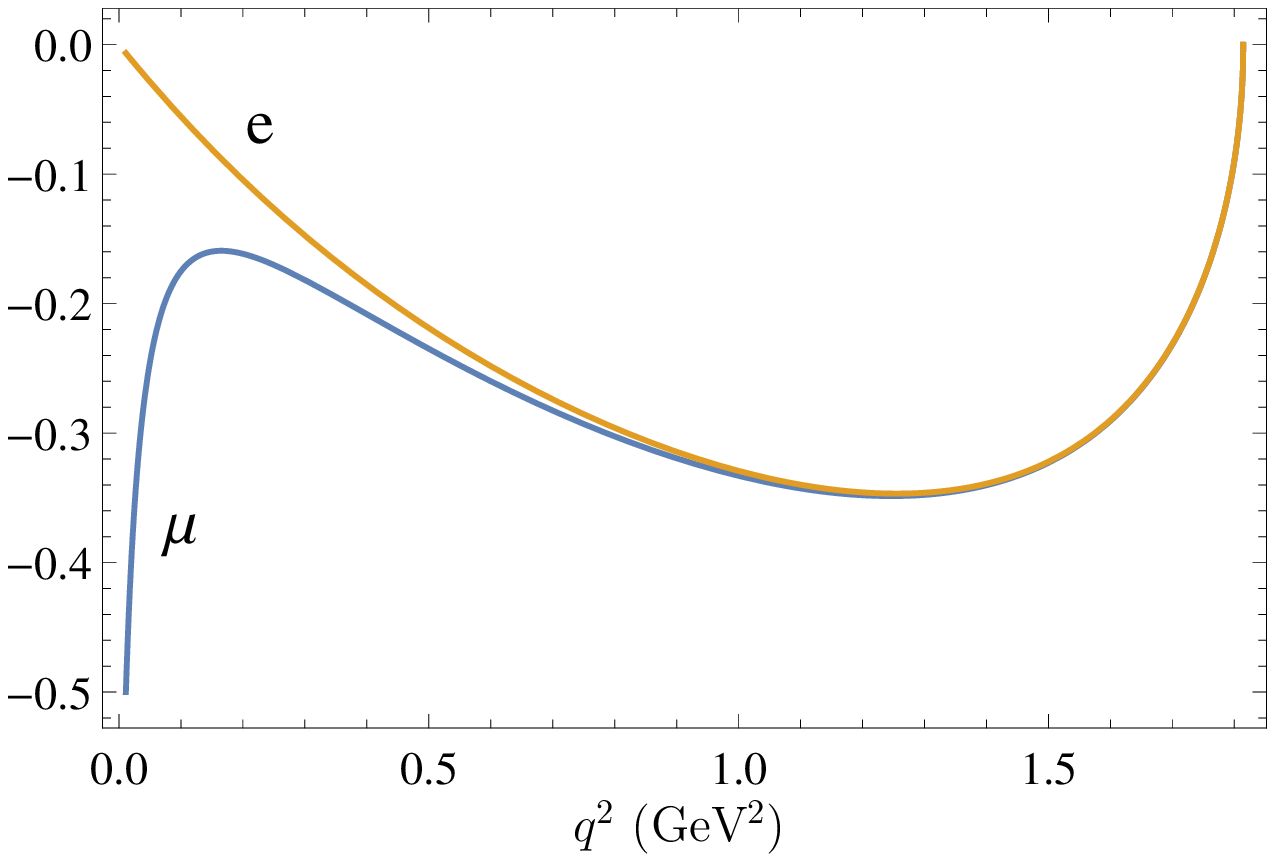}

  \caption{The forward-backward asymmetry $A_{FB}(q^2)$ in the
    $\Lambda_c^+\to \Lambda l^+\nu_l$ (left) and $\Lambda_c^+\to nl^+\nu_l$ (right)
    semileptonic decays. }
  \label{fig:afbLc}
\end{figure}

\begin{figure}
  \centering
 \includegraphics[width=8cm]{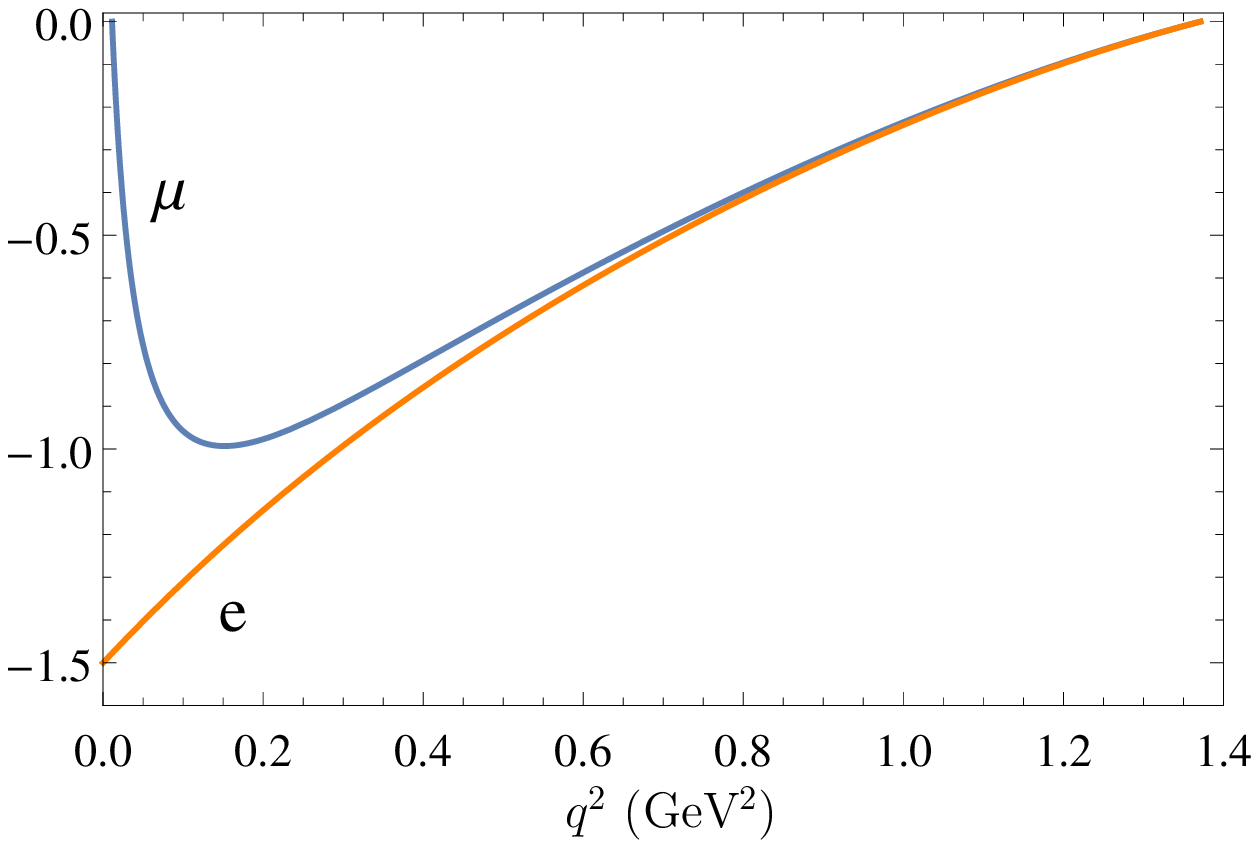}\ \
 \  \includegraphics[width=8cm]{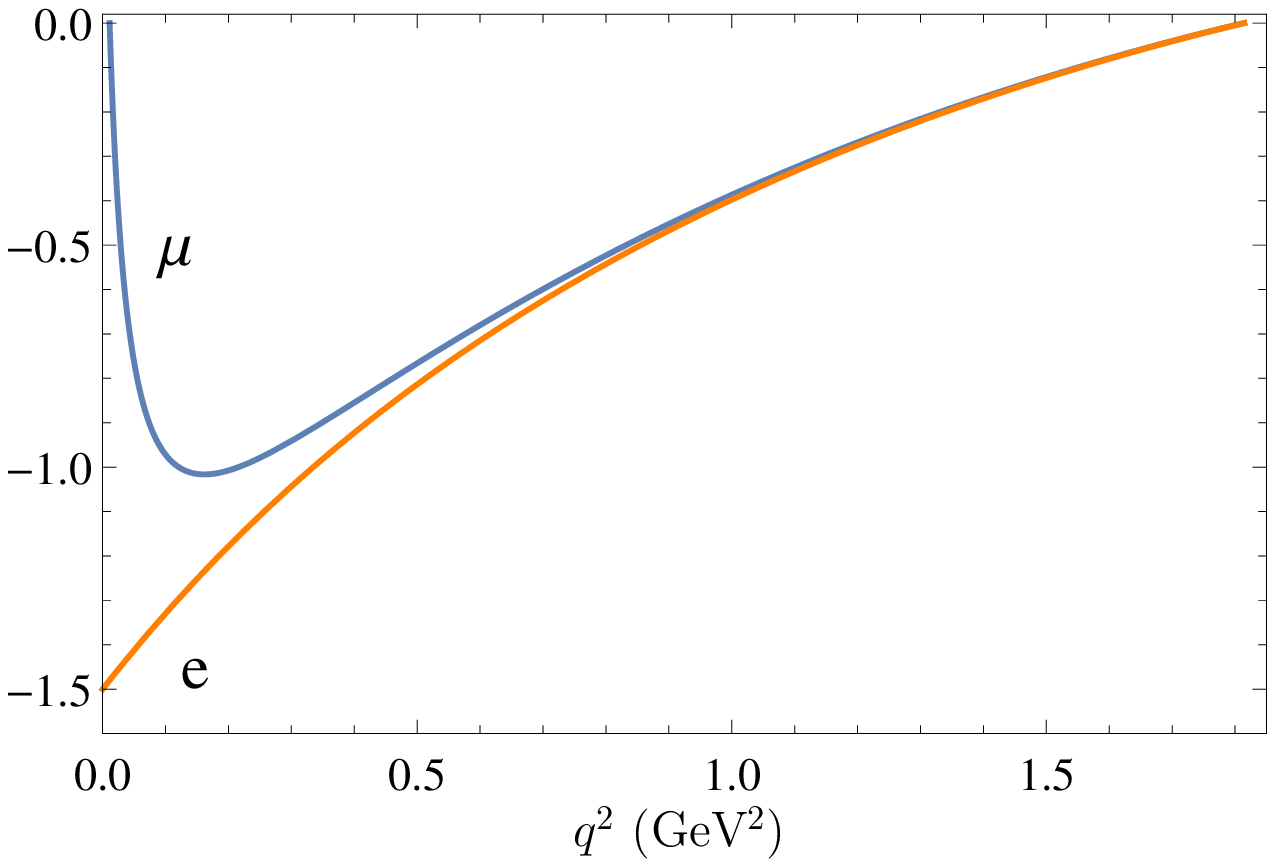}

  \caption{The convexity parameter $C_F(q^2)$ in the
    $\Lambda_c^+\to \Lambda l^+\nu_l$ (left) and $\Lambda_c^+\to nl^+\nu_l$ (right)
    semileptonic decays. }
  \label{fig:cfLc}
\end{figure}

\begin{figure}
  \centering
 \includegraphics[width=8cm]{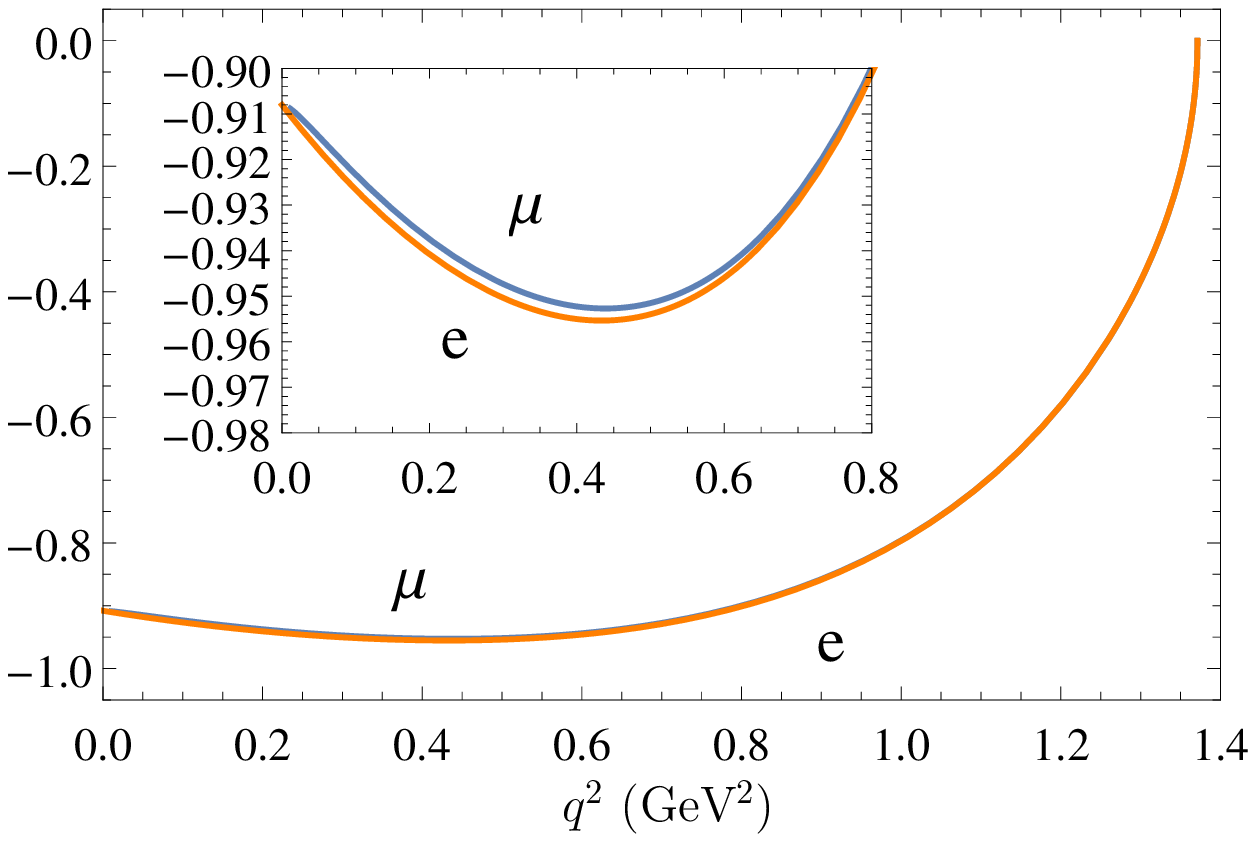}\ \
 \  \includegraphics[width=8cm]{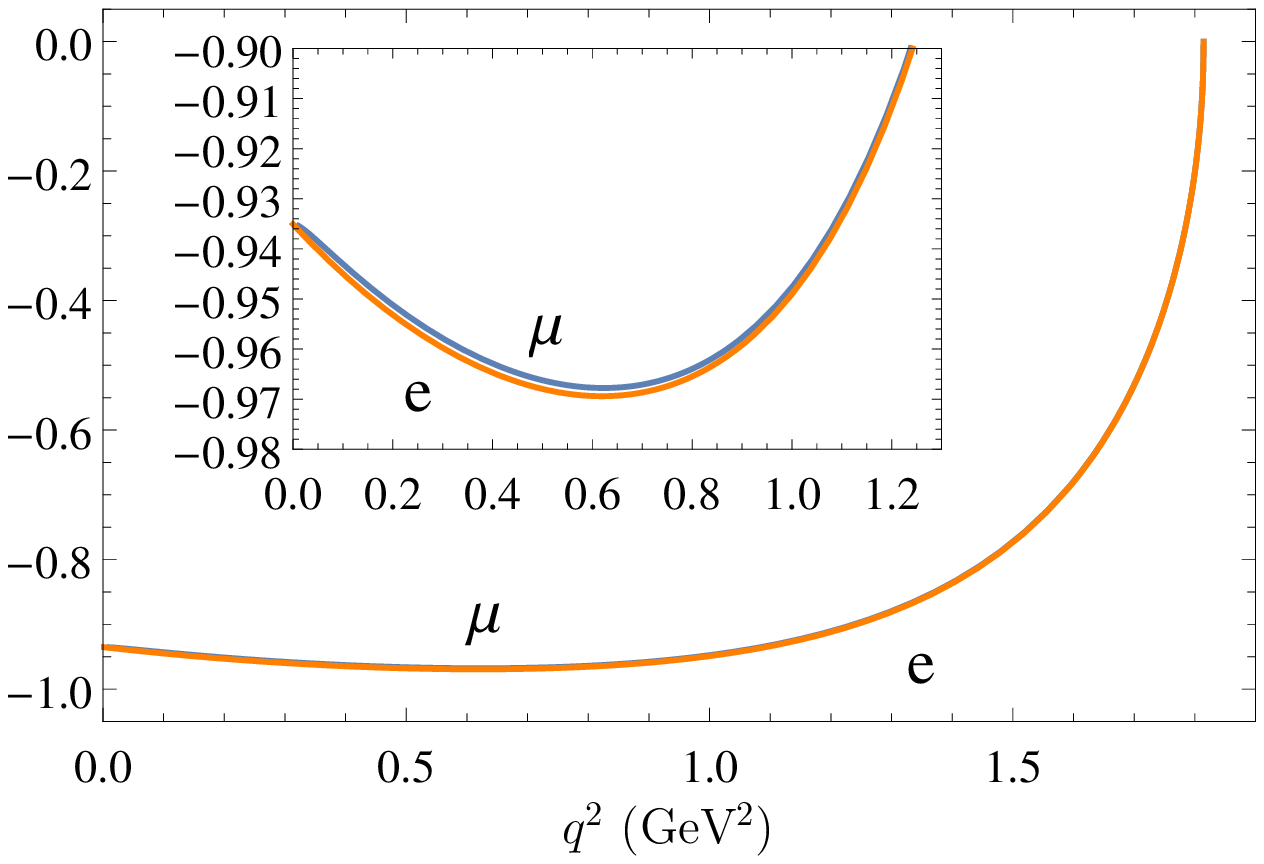}

  \caption{The longitudinal polarization $P_L(q^2)$ of the final baryon  in the
    $\Lambda_c^+\to \Lambda l^+\nu_l$ (left) and $\Lambda_c^+\to n l^+\nu_l$ (right)
    semileptonic decays. }
  \label{fig:PLLc}
\end{figure}

We plot these observables in
Figs.~\ref{fig:brLc}-\ref{fig:PLLc} both for 
$\Lambda_c\to\Lambda l\nu_l$ and  $\Lambda_c\to n l\nu_l$ ($l=e,\mu$)
semileptonic decays. From these figures we see that the plots for
decays involving electron and muon almost coincide near the zero recoil
point $q^2=q^2_{\rm max}\equiv(M_{\Lambda_c}-M_{\Lambda(n)})^2$, while the differential decay rates, the
forward-backward asymmetry $A_{FB}$ and convexity parameter $C_F$ have
significantly different behavior near maximum recoil $q^2=q^2_{\rm min}\approx0$ ($q^2_{\rm min}=0$ for electron, which we consider to be massless, and $q^2_{\rm min}=m_\mu^2$ for muon). Thus the forward-backward asymmetry $A_{FB}(q^2)$ for $q^2\to q^2_{\rm min}$ is going to 0 for the decay $\Lambda_c\to\Lambda(n) e\nu_e$ and to $-0.5$ for the $\Lambda_c\to\Lambda(n) \mu\nu_\mu$, while the convexity parameter $C_F(q^2)$ for $q^2\to q^2_{\rm min}$ is going to $-1.5$ and 0, respectively.  On the
other hand the
plots for the longitudinal polarization $P_L$ of the final baryon are
almost indistinguishable in the whole kinematical range.

The total decay rates and branching fractions are
obtained by integrating  the differential decay rates
(\ref{eq:dgamma}) over accessible kinematical region of the momentum transfer $q^2$. We
present the obtained results in Table~\ref{pred} together with the
average values of the forward-backward asymmetry of the charged lepton
$\langle A_{FB}\rangle$, the convexity parameter $\langle C_F\rangle$
and the longitudinal polarization of the final baryon $\langle
P_L\rangle$.  The quantities $\langle A_{FB}\rangle$, $\langle
C_F\rangle$ and $\langle P_L\rangle$ are calculated by separately
integrating the numerators and denominators in (\ref{eq:afb})-(\ref{eq:pl}) over $q^2$. We estimate
the errors of our calculations of the decay rates and branching
fractions divided by the square of the corresponding CKM matrix
element  $|V_{cq}|^2$, to be about 10\%. For the calculation of the
absolute values of the branching fractions we use the average experimental values of the CKM
matrix elements $|V_{cs}|=0.986 \pm 0.016$ and
$|V_{cd}|=0.225 \pm 0.008$  \cite{pdg}.

\begin{table}
\caption{Baryon decay rates, branching fractions and
  asymmetry parameters. }
\label{pred}
\begin{ruledtabular}
\begin{tabular}{cccccccc}
Decay& $\Gamma$ (ns$^{-1}$) & $\Gamma/|V_{cq}|^2$ (ps$^{-1}$)& $Br$ (\%)&$Br/|V_{cq}|^2$
  & $\langle A_{FB}\rangle$ &$\langle C_F\rangle$& $\langle P_L\rangle$\\
\hline
$\Lambda_c^+\to\Lambda e^+\nu_e$ &162&0.167 & 3.25 & 0.033 & $-0.209$&$-0.65$&$-0.86$\\
$\Lambda_c^+\to\Lambda \mu^+\nu_\mu$&157  &0.162 & 3.14 & 0.032& $-0.242$&
                                                                    $-0.55$&$-0.86$\\
$\Lambda_c^+\to n e^+\nu_e$ &13.4&0.265 & 0.268 & 0.053 & $-0.251$&$-0.60$&$-0.91$\\
$\Lambda_c^+\to n \mu^+\nu_\mu$&13.1  &0.260 & 0.262 & 0.052& $-0.276$& $-0.52$&$-0.90$\\
\end{tabular}
\end{ruledtabular}
\end{table}

In Table~\ref{comp} we compare our results for various $\Lambda_c$
semileptonic decay observables with the recent predictions of other
theoretical approaches \cite{gikls,prc,lhw,abss,lwy} (references to previous results can be found e.g. in Ref.~\cite{gikls}) and
available experimental data \cite{pdg,besiii}. The most detailed
predictions are given in the framework of the  covariant
confined  quark model  \cite{gikls}. The authors present
the values not only of decay rates and branching fractions but also
of different asymmetries and polarization parameters. We find good
agreement with their results.
The semirelativistic quark model is used in Ref.~\cite{prc}. Different
versions of QCD light-cone sum rules are employed in
Refs.~\cite{lhw,abss}. The predictions of Ref.~\cite{lwy} are based on
the application of the flavor SU(3) symmetry to $\Lambda_c$ decays,
where experimental data for $\Lambda_c\to \Lambda e\nu_e$ is taken as
an input. We
compare theoretical predictions with experimental values given in PDG
\cite{pdg} and with recent BESIII data \cite{besiii}, which are
available for the branching fractions of $\Lambda_c\to \Lambda l\nu_l$
decays. All theoretical predictions reasonably agree with data. For
the $\Lambda_c\to nl\nu_l$ no data are available at present. Most of the
theoretical predictions \cite{gikls,prc,lhw,lwy}, except the
light cone QCD sum rule approach  \cite{abss} (which predicts
significantly different values of decay form factors, see
Table~\ref{compbpiff}), give close values for the branching fractions $Br(\Lambda_c\to nl\nu_l)=0.2-0.3\%$.

\begin{table}
\caption{Theoretical predictions for the $\Lambda_c$ baryon semileptonic decay
  parameters and available experimental data. }
\label{comp}
\begin{ruledtabular}
\begin{tabular}{ccccccccc}
Parameter&\multicolumn{6}{c}{Theory}&\multicolumn{2}{c}{Experiment}\\
\cline{2-7} \cline{8-9}
& this paper & \cite{gikls} & \cite{prc}&\cite{lhw}&\cite{abss}
&\cite{lwy}  &PDG \cite{pdg}& BESIII \cite{besiii}\\
\hline
$\Lambda_c^+\to\Lambda e^+\nu_e$\\
 $\Gamma$ (ns$^{-1}$)&162&139 &236\\
$Br$ (\%)& 3.25& 2.78&4.72 &$3.0\pm0.3$& &&$2.9\pm0.5$&$3.63\pm0.43$\\
$\langle A_{FB}\rangle$ &$-0.209$&$-0.21$\\
$\langle C_F\rangle$& $-0.65$& $-0.62$\\
$\langle P_L\rangle$& $-0.86$& $-0.87$\\
\hline
$\Lambda_c^+\to\Lambda \mu^+\nu_\mu$\\
 $\Gamma$ (ns$^{-1}$)&157&135 &236\\
$Br$ (\%)& 3.14& 2.69&4.72 &$3.0\pm0.3$ &&&$2.7\pm0.6$\\
$\langle A_{FB}\rangle$ &$-0.242$&$-0.24$\\
$\langle C_F\rangle$& $-0.55$& $-0.54$\\
$\langle P_L\rangle$& $-0.86$& $-0.87$\\
\hline
$\Lambda_c^+\to n e^+\nu_e$\\
 $\Gamma$ (ns$^{-1}$)&13.4& &13.5\\
$\frac{\Gamma}{|V_{cd}|^2}$ (ps$^{-1}$)&0.265&0.20 &&&$8.21\pm2.80$\\
$Br$ (\%)& 0.268& 0.207&0.27&&$8.69\pm2.89$&$0.293\pm0.034$ \\
$\langle A_{FB}\rangle$ &$-0.251$&$-0.236$\\
\hline
$\Lambda_c^+\to n \mu^+\nu_\mu$\\
$\frac{\Gamma}{|V_{cd}|^2}$ (ps$^{-1}$)&0.260&0.19& &&$8.3\pm2.85$\\
$Br$ (\%)& 0.262& 0.202& &&$8.78\pm2.89$\\
$\langle A_{FB}\rangle$ &$-0.276$&$-0.260$\\
\end{tabular}
\end{ruledtabular}
\end{table}

\section{Conclusions}

In this paper we studied the $\Lambda_c$ semileptonic decays in the
framework of the relativistic quark model based on the quasipotential
approach and QCD. The decay form factors were  calculated in
the whole accessible kinematical range without additional model
assumptions and extrapolations. The relativistic effects were
consistently taken into account including wave function
transformations from the rest to moving reference frame and contributions
of the intermediate negative-energy states. They were expressed
through the overlap integrals of the baryon wave functions, which are
known from the  mass spectrum calculations. Such
self consistent approach significantly improves the reliability of the
obtained results. Further improvements can be achieved by considering
deviations from the quark-diquark picture of the baryons.

Using the helicity formalism and calculated form
factors we got detailed predictions for the differential and total
$\Lambda_c\to\Lambda l\nu_l$ and $\Lambda_c\to\Lambda n\nu_l$ decay
rates as well as asymmetry and polarization parameters. The obtained
results agree well with most of the previous ones
\cite{gikls,prc,lhw,lwy} and available experimental data. Our
model predicts currently unmeasured branching fraction of the
semileptonic $\Lambda_c\to nl\nu_l$ decay to be  $(0.27\pm0.03)\%$. 

\acknowledgements
We are grateful to A. Ali, D. Ebert, M. Ivanov, J. K\"orner
V. Lyubovitskij  and V. Matveev for valuable  discussions and support.

\end{document}